\documentclass[sigconf]{acmart}
\def\BibTeX{{\rm B\kern-.05em{\sc i\kern-.025em b}\kern-.08emT\kern-.1667em\lower.7ex\hbox{E}\kern-.125emX}}
    
\usepackage{nicefrac}
\usepackage{siunitx}
\usepackage{array,framed}
\usepackage{booktabs}
\usepackage{
  color,
  float,
  epsfig,
  wrapfig,
  graphics,
  graphicx,
  subcaption
}
\usepackage{textcomp}
\usepackage{setspace}
\usepackage{latexsym,fancyhdr,url}
\usepackage{enumerate}
\usepackage{algorithm2e}
\usepackage{algpseudocode}
\usepackage{graphics}
\usepackage{xparse} % argument parsing -- \edist
\usepackage{xspace}
\usepackage{multirow}
\usepackage{csvsimple}
\usepackage{balance}
\usepackage{
  tikz,
  pgfplots,
  pgfplotstable
}
\usepackage{hyperref}

\usetikzlibrary{
  shapes.geometric,
  arrows,
  external,
  pgfplots.groupplots,
  matrix
}

\pgfplotsset{compat=1.9}
\usepackage{mathtools,}

\DeclareMathAlphabet{\mathcal}{OMS}{cmsy}{m}{n}
\DeclareGraphicsExtensions{%
    .png,.PNG,%
    .pdf,.PDF,%
    .jpg,.mps,.jpeg,.jbig2,.jb2,.JPG,.JPEG,.JBIG2,.JB2}

\usepackage{xparse}
\newcommand{\bnm}{\begin{newmath}}
\newcommand{\enm}{\end{newmath}}

\newcommand{\bea}{\begin{eqnarray*}}%
\newcommand{\eea}{\end{eqnarray*}}%

\newcommand{\bne}{\begin{newequation}}
\newcommand{\ene}{\end{newequation}}

\newcommand{\bal}{\begin{newalign}}
\newcommand{\eal}{\end{newalign}}

\newenvironment{newalign}{\begin{align}%
\setlength{\abovedisplayskip}{4pt}%
\setlength{\belowdisplayskip}{4pt}%
\setlength{\abovedisplayshortskip}{6pt}%
\setlength{\belowdisplayshortskip}{6pt} }{\end{align}}

\newenvironment{newmath}{\begin{displaymath}%
\setlength{\abovedisplayskip}{4pt}%
\setlength{\belowdisplayskip}{4pt}%
\setlength{\abovedisplayshortskip}{6pt}%
\setlength{\belowdisplayshortskip}{6pt} }{\end{displaymath}}

\newenvironment{newequation}{\begin{equation}%
\setlength{\abovedisplayskip}{4pt}%
\setlength{\belowdisplayskip}{4pt}%
\setlength{\abovedisplayshortskip}{6pt}%
\setlength{\belowdisplayshortskip}{6pt} }{\end{equation}}

\newcounter{ctr}

\newcounter{mytable}
\def\mytable{\begin{centering}\refstepcounter{mytable}}
\def\endmytable{\end{centering}}

\newcounter{myfig}
\def\myfig{\begin{centering}\refstepcounter{myfig}}
\def\endmyfig{\end{centering}}

\newlength{\saveparindent}
\newlength{\saveparskip}
\newcommand{\E}{{\rm I\kern-.3em E}}

\renewcommand{\eqref}[1]{\mbox{Equation~(\ref{#1})}}

\def \part {part}

\renewcommand{\paragraph}[1]{\vspace*{6pt}\noindent\textbf{#1}\;}

\def \blackslug{\hbox{\hskip 1pt \vrule width 4pt height 8pt
    depth 1.5pt \hskip 1pt}}
\def \qed{\quad\blackslug\lower 8.5pt\null\par}
\newcounter{mynote}[section]

\newcommand\ignore[1]{}

\newcounter{rcnote}[section]

\newcounter{mrnote}[section]

\newcounter{fknote}[section]

\newcounter{anote}[section]

\DeclareMathSymbol{\mlq}{\mathord}{operators}{``}
\DeclareMathSymbol{\mrq}{\mathord}{operators}{`'}

\newcommand{\rhf}[2]{R_{f, \gamma}}

\DeclareDocumentCommand{\edist}{o o}{
  \ensuremath{
    \IfNoValueTF{#1}{{d}}{{\sf d}(#1,#2)}
  }
}

\newcommand{\olrk}[1]{\ifx\nursymbol#1\else\!\!\mskip4.5mu plus 0.5mu\left(\mskip0.5mu plus0.5mu #1\mskip1.5mu plus0.5mu \right)\fi}

\NewDocumentCommand{\indseq}{ O{1} O{r} }{{#1}\ldots {#2}}

\begin{document}
%\fontfamily{lmr}\selectfont
% \def\thetitle{A Practical Way to Generate Strong Keys from Noisy Data}
\fancyhead{}
\def\thetitle{The Model Mastery Lifecycle: A Framework for Designing Human-AI Interaction}
\title{\thetitle}

\author{Mark Chignell}
\affiliation{%
  \institution{University of Toronto}
  \city{Toronto}
  \country{Canada}
}
\email{chignell@mie.utoronto.ca}

\author{Mu-Huan Chung}
\affiliation{%
  \institution{University of Toronto}
  \city{Toronto}
  \country{Canada}
}
\email{mhm.chung@mail.utoronto.ca}

\author{Jaturong Kongmanee}
\affiliation{%
  \institution{University of Toronto}
  \city{Toronto}
  \country{Canada}
}
\email{jaturong.kongmanee@mail.utoronto.ca}

\author{Khilan Jerath}
\affiliation{%
  \institution{Sun Life Financial Inc.}
  \city{Toronto}
  \country{Canada}
}
\email{khilan.jerath@sunlife.com}

\author{Abhay Raman}
\affiliation{%
  \institution{Sun Life Financial Inc.}
  \city{Toronto}
  \country{Canada}
}
\email{abhay.raman@sunlife.com}

\date{}

\begin{abstract}
  The utilization of AI in an increasing number of fields is the latest iteration of a long process, where machines and systems have been replacing humans, or changing the roles that they play, in various tasks. Although humans are often resistant to technological innovation, especially in workplaces, there is a general trend towards increasing automation, and more recently, AI. AI is now capable of carrying out, or assisting with, many tasks that used to be regarded as exclusively requiring human expertise. In this paper we consider the case of tasks that could be performed either by human experts or by AI and locate them on a continuum running from exclusively human task performance at one end to AI autonomy on the other, with a variety of forms of human-AI interaction between those extremes. Implementation of AI is constrained by the context of the systems and workflows that it will be embedded within. There is an urgent need for methods to determine how AI should be used in different situations and to develop appropriate methods of human-AI interaction so that humans and AI can work together effectively to perform tasks. In response to the evolving landscape of AI progress and increasing mastery, we introduce an AI Mastery Lifecycle framework and discuss its implications for human-AI interaction. The framework provides guidance on human-AI task allocation and how human-AI interfaces need to adapt to improvements in AI task performance over time. Within the framework we identify a zone of uncertainty where the issues of human-AI task allocation and user interface design are likely to be most challenging. 

\end{abstract}

\begingroup
\mathchardef\UrlBreakPenalty=10000
\maketitle

% Section I
\section{Introduction}
\label{sec:intro}

We live in an age of model-based prediction and classification. A chess playing model, for instance, seeks to predict what the best move should be in each situation encountered during a chess game \cite{campbell2002deep}. In recent decades models have begun to outperform humans in chess as well as many other games and tasks. In chess, an expert player is referred to as a master. Humans have always been familiar with the idea of mastery. In many occupations, such as blacksmithing, masters would pass on their skills to apprentices who might, with the passage of time, become more skillful than the masters who taught them. The same process of mastery can be observed in a wide range of occupations and pursuits including artistic, sporting, and technical endeavours. The concept of mastery is a common way of recognizing human expertise in many fields. Thus people in professions such as the law, engineering and medicine have to go through a training and licensing process before they can practice in their chosen profession. 

One problem with AI models is that the necessary licensing and certification processes are not yet in place. There has been a focus on making AI models increasingly smart, but with AI capable of performing more and more tasks, the problem of AI governance becomes increasingly salient. We need to ensure that AI works in the service of human needs and values \cite{shneiderman2022human}, rather than as a competing form of intelligence. We have not reached the stage where AI models or agents can be licensed as radiologists or pharmacists, for example, yet AI techniques are being applied to an ever-expanding set of tasks and the issues of how to implement newly available AI expertise into task workflows is becoming increasingly pressing. When AI models can beat the best human players in complex games like chess and Go, and can perform many tasks autonomously, the issues of AI governance, licensing, and human-AI coordination become unavoidable. 

As with human-human interactions, different forms of human-AI interaction (HAII) can be distinguished \cite{chignell2023evolution}. AI can be “in charge” (or autonomous), it can be an equal partner, where decision making is made jointly and collaboratively by human and AI agents, or it may be subservient, carrying out the requests made by humans who act as supervisory controllers \cite{xu2024applying, yildirim2023investigating}. In this paper we examine when and how AI should be in charge, and how we should adapt to increasing levels of progress towards AI mastery. Our suggestions relating to the integration of AI into tasks and workflows supplement approaches that argue for human priority \cite{sellen2023rise}, with technology that is human-centred, controlled and managed \cite{shneiderman2022human}. 

Another factor that complicates AI implementation is resistance to recognizing AI mastery \cite{dawes1974linear} when it occurs. This may be partly due to a Dunning-Kruger effect \cite{kruger1999unskilled} where experts may overestimate their relative decision making ability as compared to AI models. Ideally, decisions on how to use AI models that are approaching mastery should focus on a performance-based assessment of decision making outcomes, while recognizing the need to preserve human values \cite{lai2023towards}. 

The concept of model autonomy (as in the goal of autonomous driving) is well understood, but autonomy is a state that should occur only after mastery has been achieved. A major challenge for human-AI interaction researchers in the future will be to better understand the process of progressing towards model mastery and the human interface design requirements at each stage of progression towards model mastery. Considerations of model mastery build on, and complement existing approaches to HAII. 

One aspect of previous approaches to HAII has involved recognizing and designing for the different skills and abilities that AI models and humans have, while also designing user interfaces that reduce the communication mismatches between the respective inputs and outputs of human and AI agents. Amershi et al. \cite{amershi2019guidelines} provided foundational guidelines for HAII, emphasizing the importance of understanding and enhancing the interactions between humans and AI systems across various domains. Similarly, Li et al. \cite{li2023assessing} demonstrated how human preferences can be gauged early in product development, highlighting the human ability to discern AI performance levels, which can guide the implementation of AI systems. Gama et al. \cite{gama2022implementation} reviewed frameworks for AI integration in healthcare, suggesting a broader applicability and adaptation of these frameworks to effectively incorporate AI technologies. However, while these guidelines offer a robust foundation for designing human-AI systems, they do not fully address the complexities of AI mastery, and the changing nature of human and AI roles as AI progresses towards mastery for a particular task. 

There are a number of impediments to recognizing and adapting to a state of “model mastery” (where AI agents/models consistently outperform human experts). Ideally, in a situation where model mastery exists, the application of the relevant model would be deemed essential in performing a task. There is a wealth of work since the 1950s showing that models often outperform the experts that trained them, yet acceptance of model mastery tends to lag behind the actual performance of models \cite{meehl1954clinical,dawes1974linear,dawes2005ethical}. 

We also need to consider cases where there is useful, but incomplete, progress towards AI mastery. If AI models provide useful capability, but have not achieved mastery, then we face the problem of how to successfully integrate them into a team and how to deal with their shortcomings, in the same way that successful companies get the best out of employees who have a mixture of useful skills and shortcomings. As AI agents or models transition from being only marginally useful to being able to handle particular tasks autonomously, we need to develop suitable interfaces to handle those intermediate cases where AI models are useful but sometimes brittle. We also need to develop procedures for recognizing when AI mastery has occurred for a task, and to reconfigure sociotechnical systems accordingly so that AI mastery can be safely integrated into general work practices and societal systems. 

Mastery occurs when AI agents are able to surpass the intelligence/skill of their human tutors in a process that many find increasingly disturbing (e.g., the singularity envisaged by Ray Kurzweil \cite{kurzweil2000age}, and the concerns raised by Geoffrey Hinton \cite{UofTNews}). The concept of AI mastery is very broad, and may be difficult to assess in some domains. To what extent, for instance, do the sometimes surprising outputs of large language models (LLMs) \cite{elkins2020can} indicate a high level of intellect and mastery? As expectations of AI mastery increase there is the risk that benefits observed with the introduction of AI may sometimes be at least partly due to placebo effects where people perform better when they think that they are being helped by AI, even if it is not actually assisting them \cite{kosch2023placebo}. 

The possible roles of AI in a task can be illustrated with an analogy. Say that the manager of a team has hired someone to work on important new functionality in the company’s inventory management system. If the new hire is a highly skilled developer, who has successfully implemented similar software in the past, the manager may be willing to let the developer work on the task relatively autonomously, reporting back when the work is almost done. However if the skills and experience of the new hire are not as well suited to the required task, then the manager may supervise the work more actively, or may incorporate the new hire into a team of developers who work on the task. More resources and more opportunities to check on the work being done may be beneficial, but they come at the cost of more communication effort as people need to coordinate the work, check on progress, etc. This analogy is apt, because the AI model or agent is like the new hire, with decisions needing to be made about how it performs its work. Will it work relatively autonomously, or if not, how will it coordinate and communicate with the human agents who are also performing the work? 

In the remainder of this paper we make the following contributions. 

\begin{itemize}
    \item The construct of model mastery is introduced and is related to past work comparing human and algorithmic expertise
    \item A framework is presented for human-AI task allocation at different levels of progress towards model mastery  
    \item The Model Mastery Lifecycle is introduced 
\end{itemize}

%%% Local Variables:
%%% mode: latex
%%% TeX-master: "main"
%%% End:

%  LocalWords:  biometrics cryptographic parallelized lossy
    % basic introduction
\section{Model Mastery}
\label{sec:mastery}
Models become competitive with humans in performing tasks when they outperform the practitioners in a field (e.g., doctors making diagnostic decisions), and they achieve mastery when their superiority to human experts performing a particular task is undisputed. For instance, AI chess programs have achieved model mastery and the best human chess players in the world are consistently beaten by them. It is easy to demonstrate mastery in a game like chess, where ratings are based on the quality of player that a human or AI agent can beat. In other areas the measurement of mastery may be more controversial. For instance, an AI system that makes diagnoses or predicts health outcomes may have to deal with a wide range of patients and may sometimes make errors, thus making it harder to demonstrate that the model’s decision making is overall superior to the human experts currently doing the task \cite{oberije2014prospective}. 

Early demonstrations of model mastery using simple statistical models (e.g., linear regression) had relatively little impact on practice. However, the second decade of the twenty-first century saw increasing acceptance of machine learning and AI approaches in fields such as healthcare \cite{shuaib2020increasing,thompson2018artificial,froomkin2019ais}, with models exceeding human performance in a number of domains.

A key reason why even simple models often outperform well-trained human experts is their consistency. Unlike humans, who can vary greatly in their predictions due to factors like mood, illness, cognitive biases \cite{tversky1974judgment}, and sometimes forgetfulness, models maintain a uniform approach in their predictions. This raises important questions about the future role of algorithms in decision-making, suggesting that any shift towards more autonomous systems will need to be supported by careful management of trust and oversight in these technologies.

Figure \ref{fig:1} provides a schematic view where a potentially complex multidimensional space of outcomes has been compressed into single dimensions (one for predicted, and the other for actual, outcomes) for the purpose of illustrating why a consistent model outperforms an expert whose predictions are more variable. In a task where quality of performance is assessed in terms of how well predictions match outcomes, ideal performance would be shown as a diagonal line on the figure with predictions (e.g., of a numerical quantity) matching outcomes for all instances. In practice both model and human experts will generate imperfect predictions because of imperfect knowledge, and extra variables that they do not account for. However, the human predictions will generally be more variable as shown by the points in the figure coloured green (vs. orange). Thus, due to the greater variability in cognition and behaviour, expert predictions will vary from the ideal to a greater extent than model predictions. In situations where much of the variation involves deviations from the normal decision making process due to implicit (and accurate) knowledge that the human has (e.g, about edge cases), the variation will be beneficial and the model will continue to serve as an apprentice. However, if the preponderance of human variation is due to inconsistency and does not improve decisions, then the model will show mastery. 

\begin{figure}
    \centering
    \includegraphics[width=\linewidth]{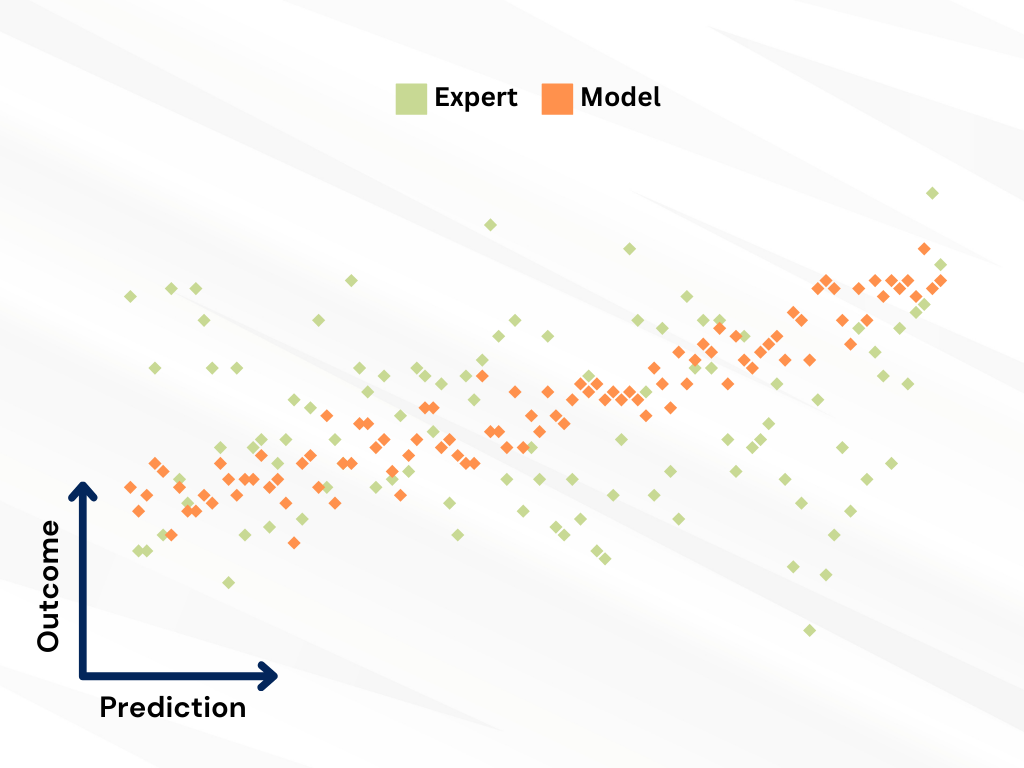}
    \caption{A simple (2D) illustrative example of potential model mastery, showing how human judgments show greater variation around a line of estimated best fit}
    \label{fig:1}
\end{figure}

Model mastery is easy to demonstrate in situations where there are not enough human experts to carry out a task on a large scale, or there is too much data for them to take account of. An example of this kind of task is screening. In recent years it has become customary for machine learning models to outperform human screeners (e.g., recognizing faces of incoming passengers at airports) and this first type of model competitiveness may be regarded as unsurprising, as the models typically have access to large amounts of data (e.g., databases of face images) that can be easily processed computationally, but that are difficult for humans to deal with. As further examples, language understanding, machine translation, and question answering are areas where the problem is too vast for an apprenticeship style of training and where algorithms are used, along with vast amounts of data and text, to infer the knowledge needed. 

Traditional ML evaluation metrics such as AUROC (Area under the Receiver Operating Characteristic Curve) and F1 (a weighted combination of precision and recall) can be used to compare model and human performance to quantify the degree of model mastery, but the problem here is that not all errors are created equal, and thus controversy over the relative merits of human and AI performance may still remain in spite of evaluations based on ground truth data, because the costs of different kinds of errors are controversial and humans and AI models tend to be differentially susceptible to different types of error. 

AI superiority, often attributed to its vast data access and computing power, is less surprising than its ability to outperform human experts even in domains where it has been trained by those experts \cite{chung2023implementing}. There is a kind of paradox where experts may recognize model mastery too late, while more naive users may assume model mastery too early. An example of this latter situation is "algorithm appreciation" \cite{logg2019algorithm} where non-experts sometimes trust algorithms more than human advice even if they are not provided with evidence of model mastery. 
%%%%%%%%%%%%%%%%%%%%%%%%%%%%%%%%%%%%%%%%%%%%%%%%%%%%%%%%%%%%%%%%%%%%%%%%%%%%%%%%
\section{Adaptation of Human-AI Interaction to Increasing Model Mastery}
\label{sec:adaptation}

Human-human interaction is a complex topic that is relevant to a wide range of fields (each with their own theories and approaches) including the law, education, business, healthcare, psychology and sociology. Adding AI to the mix increases this complexity even more \cite{mahmud2023study,cai2019hello}. People with varying levels of expertise (relative to the tasks being performed) may be interacting with AI models (also with varying levels of expertise on different types of task), as can be seen in Figure \ref{fig:2} below. 

In cases where human users are experts, we can consider cases where the model is error prone vs. high performing (top two quadrants of Figure \ref{fig:2}). For cases where the model is error prone, experts can train the model (typically using active learning \cite{settles2011theories} to ensure that the training process is efficient). For high performing models where users are experts there is a choice to be made. If model mastery is accepted, then the model may be provided with a high degree of autonomy, with ultimate autonomy being automatic implementation of the AI model’s decisions. However, if the model mastery is not fully accepted then a supervisory control approach can be used, where one or more experts oversee the model performance and may make the final decision on whether to accept the AI model’s advice or not. 

For common (non-expert) users (the situation in the lower two quadrants of Figure \ref{fig:2}), supervisory control will generally be problematic because the user may not have enough expertise to make a suitable judgement of whether or not the model’s advice is acceptable or should be acted upon. If the model is high performing but users are relatively unskilled (lower right quadrant of Figure \ref{fig:2}), then there may be more willingness to accept model mastery. When both AI models and humans are unskilled (lower left quadrant of Figure \ref{fig:2}) crowdsourcing may be a workable strategy. An overall labeling process that applies consensus labels may lead to higher quality final labels than the individual user-assigned labels on which the consensus is based.  

\begin{figure}
    \centering
    \includegraphics[width=\linewidth]{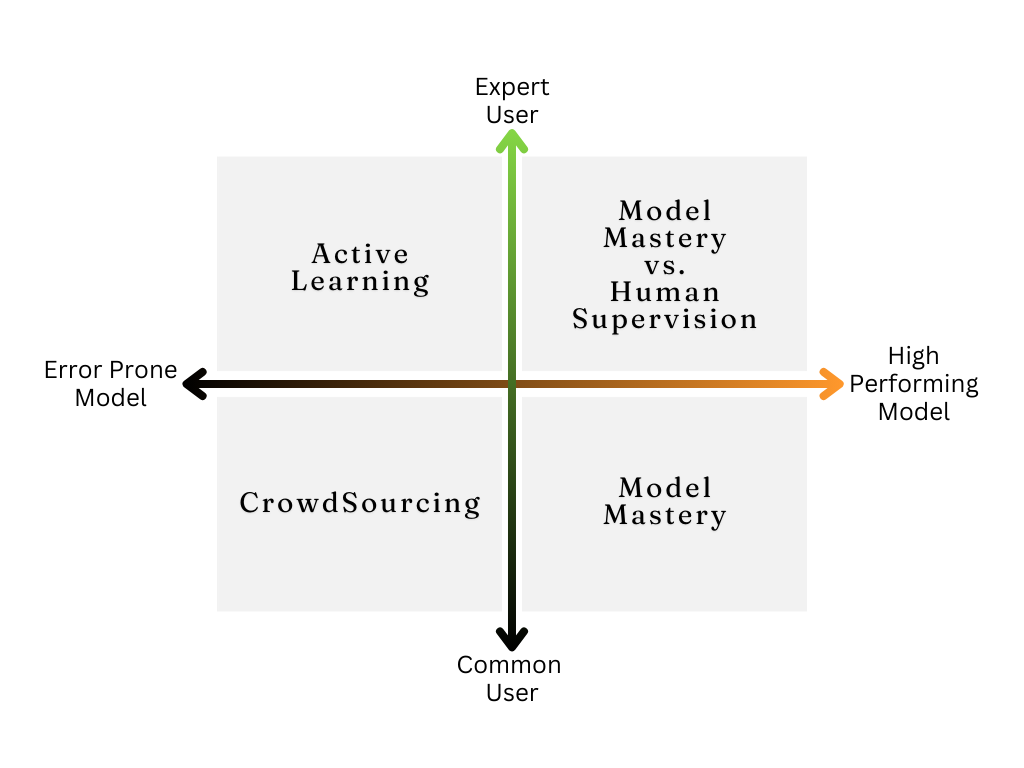}
    \caption{The design space for HAII, showing four quadrants of model performance versus human expertise. Recommended HAII strategies are shown within each quadrant}
    \label{fig:2}
\end{figure}

Figure \ref{fig:2} shows a cross-sectional view of human vs. AI expertise. While individual people may transition from the lower to upper quadrants through training, the general pools of experts and users will remain in the upper and lower quadrants respectively. In contrast, AI model performance (expertise) for a particular task will generally move from left to right in the figure as the model is trained and enhanced. 

Growing AI/automation expertise changes the role of the human, while also creating new human-AI interaction challenges, with automated driving being an example of this process.  There is a strong business case for changing the role of the human and increasing the role of automation and AI in driving. Many people don’t like driving or are not particularly good at it, and as people get older and more physically and cognitively frail, autonomous driving might be able to help them maintain the mobility that they would otherwise lose. 

Hancock et al. \cite{hancock2020challenges} explored the evolution of vehicle automation from fully human-driven to completely autonomous systems. They cited the ASME model of levels of automation (LOA), from Level 0, where the driver controls all tasks, to Level 5, where no human intervention is required. Key transitional phases were highlighted \cite{hancock2020challenges}, especially Level 3 (Conditional Automation) where the driver must be ready to take over when requested, and Level 4 (High Automation) where the vehicle handles all driving tasks but may allow for human override. Figure \ref{fig:3} shows a schematic overview of the progression towards mastery in the case of automated driving.

\begin{figure}
    \centering
    \includegraphics[width=\linewidth]{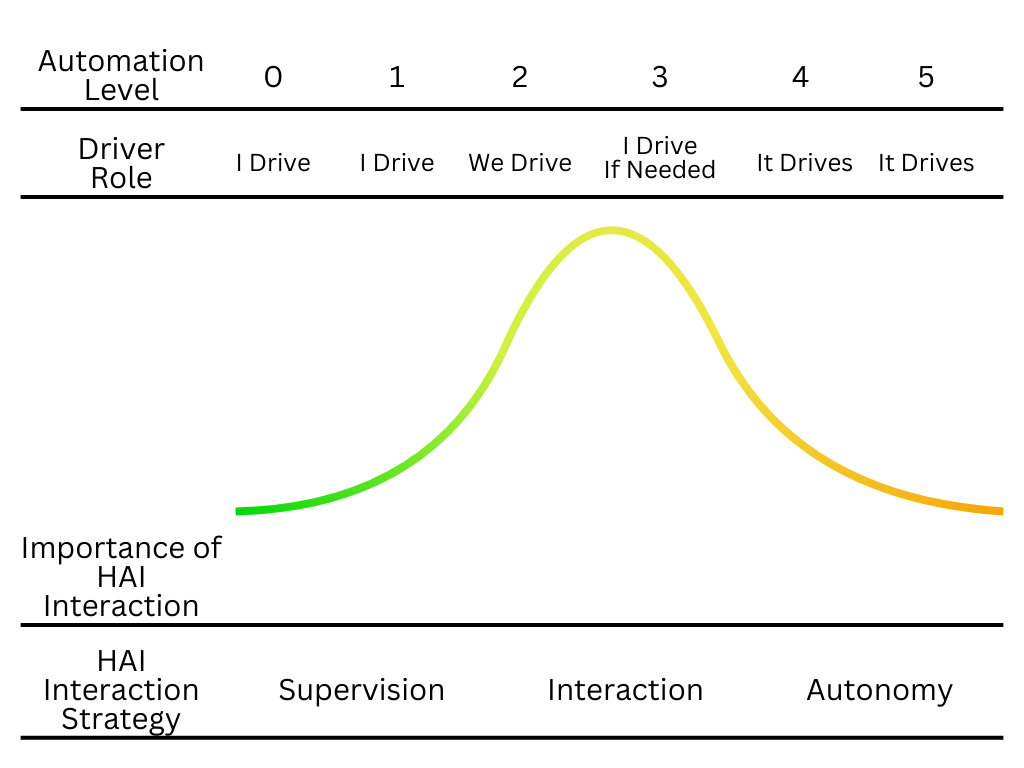}
    \caption{Schematic overview of the progression towards mastery in the case of automated driving}
    \label{fig:3}
\end{figure}

Trust and HAII are crucial concerns in increasingly automated vehicles. In lower levels (level 0-2), the driver is actively engaged and must constantly monitor vehicle operations. At mid-level automation (level 3), trust issues become prominent as drivers rely on the vehicle for full environmental monitoring but must remain prepared to intervene. At higher levels (level 4-5), the vehicle gains full control, requiring absolute trust from passengers as their role shifts completely from driver to observer. The transition to fully autonomous vehicles necessitates robust trust calibration and clear communication about the vehicle’s capabilities and limitations. Regulatory frameworks and independent third-party testing are essential to ensure safety and efficacy as these technologies are increasingly implemented in vehicles across a wider range of roadways. This example shows how the type of HAII, and related issues such as trust, changes as AI/automation progresses towards mastery.

As automated driving technology transitions from lower to higher levels of automation, the identification of who is the co-pilot \cite{sellen2023rise} is likely to change. At low levels of automation the human driver is the “pilot” and the automation is the co-pilot. However, as automation progresses to higher levels more of the driving is carried out by the automation and the human transitions to a co-pilot role. Thus, rather than automatically assigning the co-pilot task to the AI model/agent, we need to consider which agent (human or AI) is naturally in charge and then assign the co-pilot to the remaining agent (human or machine). Perhaps, in the future, the etiquette of HAII across different levels of humans vs. AI expertise will be well understood. However, at present people in general are not well-trained, or experienced, in terms of how to interact with AI models. Human ability will tend to change little over the short term, so in most tasks AI ability will be increasing relative to human ability over time. This effect is captured schematically in Figure \ref{fig:4}, where we see that the green line (human ability in a given task given an appropriate level of training) is relatively flat, whereas model performance is increasing over time.

We hypothesize that there will generally be a four-stage process, each stage requiring different types of HAII. As model performance improves, it goes from being clearly worse than a human (where if used, the model needs to be supervised), and approaches a region where the model, while still judged to be inferior in its performance, is nevertheless useful enough to be seen as a partner and interacted with. This region of interaction will require careful HAII design \cite{huang2021engaged}. A third stage is reached when model expertise has approached close enough to human capability that it may become difficult to determine the relevant tasks that should be carried out by the human vs. AI agents, and unclear as to what type of HAII should be used. In this zone of uncertainty the relative expertise of models vs. humans may be in dispute. 

HAII within the zone of uncertainty may need to be relatively fluid and more complex, as human experts attempt to calibrate model performance across different types of scenarios, and identify edge cases where different types of interaction are required. We can imagine situations where busy experts may sometimes handoff tasks (in the zone of uncertainty) to AI models/agents and AI agents may offer help in cases where their models predict that their task performance is highly likely to be accurate. While this type of task handoff back and forth between human and AI agents may seem straightforward, policies and interfaces that facilitate these ongoing task allocations and re-allocations will need to consider human psychology and preferences. For instance, Adam et al. \cite{adam2024navigating} found that when an information system (IS) autonomously offered to assume tasks without user prompts (IS-invoked delegation), users experienced a perceived threat to their self-concept, particularly when they feel a lack of control over the process. 

After the zone of uncertainty, as the model increasingly approaches mastery, model autonomy becomes an option, with fewer and fewer edge cases being identifiable. 

\begin{figure}
    \centering
    \includegraphics[width=\linewidth]{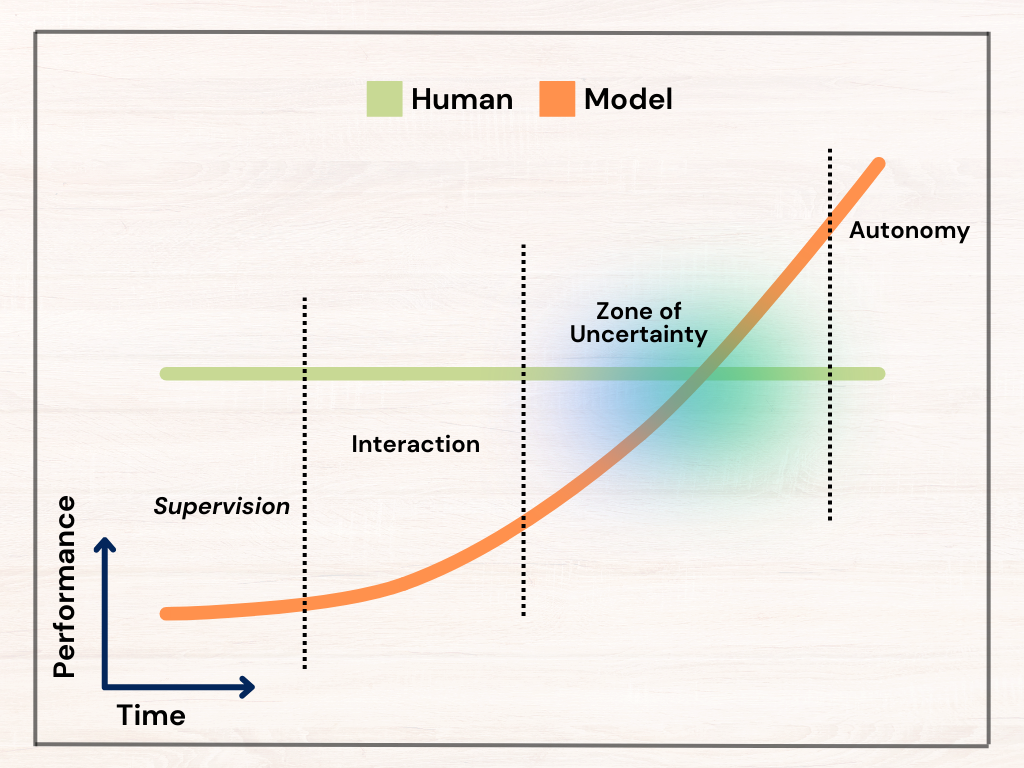}
    \caption{The Model Mastery Lifecycle: Stages in the Growth of Model Mastery showing how overall HAII strategy changes as model mastery increases}
    \label{fig:4}
\end{figure}

The growth of automation in driving, and the resulting changes to the role of the driver mirrors a process that started earlier in the field of aviation. Much of the job of pilots flying large passenger planes involves supervisory control \cite{sheridan2012human}. Aviation psychologists and human factors engineers have spent decades working on appropriate human interfaces for highly automated aircraft, motivated by the requirement to improve safety and reduce the risk of accidents. Their experience shows how challenging design of HAII is likely to be, for situations where AI models are in the intermediate stages of progression towards model mastery. The following questions posed by Sheridan illustrate the types of issue that need to be addressed in aviation, and similar questions will likely need to be asked in other domains:

\begin{itemize}
    \item Should there be certain states or a certain envelope of conditions for which the automation will simply seize control from the pilot?
    \item Should the computer deviate from a programmed flight plan automatically if critical unanticipated circumstances arise?
    \item If the pilot programs certain maneuvers ahead of time, should the aircraft execute these automatically at the designated time or location, or should the pilot be called upon to provide further concurrence or approval?
\end{itemize}

The Terminal Collision Avoidance System (TCAS) is another example from the field of aviation, where AI is being integrated at varying levels of autonomy. TCAS, developed to mitigate mid-air collisions, highlights the potential of automation to exceed human capabilities in specific, well-defined scenarios by compensating for human limitations in reacting to situations fast enough \cite{kochenderfer2012new}. 

To further illustrate the generality of the model mastery approach, and associated HAII challenges, we consider examples from the rapidly evolving domain of healthcare. There are many different tasks in this domain that are gradually transforming into human-AI interactions. The varying levels of autonomy and trust issues evident across different medical domains, reflect both the level of maturity of AI applications and the complexities of their respective fields.

In disease diagnosis, Topol et al. \cite{topol2019high} described how systems like IBM Watson leverage AI to analyze medical imaging and clinical reports with remarkable accuracy, significantly enhancing diagnostic precision and speed. This domain benefits extensively from AI’s ability to process vast amounts of data quickly, thereby supporting clinicians, optimizing health system operations, and enabling patient self-monitoring. Despite this high autonomy, there remain substantial concerns regarding biases, privacy, and transparency, necessitating a cautious approach to model implementation in practical settings.

The use of AI has also made strides in radiation oncology, primarily being used to enhance processes like image segmentation, dose optimization, and quality assurance \cite{thompson2018artificial}. Although AI shows potential to improve efficiency and accuracy in these areas, it is still under development and faces challenges related to data quality and the need for interpretability (so that the results provided by AI models can be trusted).

The progress of AI models towards mastery is slower in areas such as mental health services, where human empathy may be an important part of the interaction with patients. However, even in mental health AI systems are being used. For instance, the HAILEY system \cite{sharma2023human}, supports peer supporters by providing feedback to enhance communication in text-based support. 

The preceding brief survey of HAII contexts shows that progress towards model mastery varies between application areas. In aviation settings, and particularly in TCAS, there is a high degree of automation, which leads in some situations to almost complete autonomy. For example, the introduction of glass cockpit technology, and the automation of take-off and landing processes, has significantly transformed the role of pilots turning them into supervisory controllers in most situations. Technologies like Autoland, which can handle landings in poor visibility, demonstrate how critical aviation tasks, once exclusively managed by human pilots, are now being automated, increasing both safety and consistency in operations.

Applications such as clinical decision making in healthcare tend to be less highly automated and AI, when available, tends to have an advisory role. Every application of AI in a real world task will tend to be at some point on the model mastery lifecycle (Figure \ref{fig:4}), with model mastery tending to increase over time for different applications, but at different rates. As shown in the figure, as AI expertise improves human-AI task allocation will transition through four main stages: 

\begin{enumerate}
    \item The supervision stage where the human is clearly in charge
    \item The interaction stage where human and AI work together as partners
    \item The zone of uncertainty, where the relative strengths of human and AI expertise are unclear and where the allocation of tasks between human and AI elements may need to be negotiated, and dependent on the situation
    \item The autonomy stage where the AI can carry out the task relatively independently, possibly with a light amount of monitoring and supervision.
\end{enumerate}

In this paper we have highlighted model mastery as a general concept that will influence HAII design for particular applications. The generality of the model mastery approach should make it possible to define different styles of interaction that should be used at different stages of the model mastery lifecycle. HAII will be important during the interaction phase of the lifecycle and likely even more important in the zone of uncertainty. In that zone, defining the role of the human vs. AI systems, allocating tasks in real-time and managing task handoffs between humans and AI will present complex challenges. These challenges will occur across a range of domains, including those of aviation, driving and healthcare considered above. All HAII designs will need to provide an appropriate balance between human oversight and automated control. Another general issue is the need to calibrate trust appropriately, ensuring that automation enhances human capabilities without replacing the critical insights and judgments that only humans can provide. This ongoing balance seeks to leverage technological advancements to improve safety and efficiency while maintaining essential human involvement in critical decision-making processes.

Each application that uses AI will need to formulate its own approach to HAII, taking into account both the general requirements relating to its position in the model mastery lifecycle, as well as specific requirements reflecting the properties of the domain within which the application is situated. 

We are not yet in a position where it is possible to develop prescriptive approaches to the design of HAII, but there are general principles that are likely to be particularly important in the zone of uncertainty (Figure \ref{fig:4}), and we list a few of the most important principles below. 

\begin{enumerate}
    \item The human role should adjust appropriately as AI expertise increases in an application.
    \item There should be robust methods for ensuring the safety of human-AI systems, and alignment of human-AI system performance with human values and ethical requirements. 
    \item There should be mechanisms for assessing and adjusting the level of trustability of AI performance and for communicating that trustability to the people within the human-AI system. These should include methods for improving the transparency of AI processes (cf. \cite{skraaning2021human}).
    \item There should be an iterative approach to AI implementation, where AI systems are rolled out incrementally based on their reliability and user feedback. 
    \item Development of human-AI systems will need to accommodate differences in human skills and preferences. Depending on practices and policies, work related human-AI systems may be developed for use by highly skilled and trained users/operators, or may be designed to accommodate a more diverse use group, which will likely require the customization of HAII for different types of user or worker. 
\end{enumerate}
\section{Inertia in Accepting Model Mastery}
\label{sec:inertia}

Imagine that a society had a privileged group of people who were assigned tasks, not because they were good at those tasks, but because of their status. Wars would be fought by incompetent generals, operations would be carried out by incompetent surgeons, and banks would fail because of the incompetence of their executives. Clearly this society would be less successful than one where the most competent people were put in charge of key tasks. As models achieve mastery, failure to fully implement them may be costly. However, experience with clinical decision making have shown that experts may be resistant to accepting model mastery \cite{grove1996comparative} and may base their resistance on a number of arguments that have varying degrees of legitimacy. One argument can be characterized as the “good people on both sides argument”. 

The argument here is that the model might work well in some cases but that the expert should still be able to be involved in the decision and over-rule the model in the cases where the expert “knows” that the model is wrong. Sometimes this argument might be effective. For instance, human advice may usefully overrule a model that is brittle, and the expert can provide a better decision in an edge case that the model is unprepared for. However, if experts get to pick and choose when they can intervene, there is a danger that in many cases overall decision making outcomes may be worse, since experts may not really know when they are in a better position than the model to make the decision. 

Another argument (in a clinical setting) can be characterized as “models may be good at dealing with cases in general, but I know my patients better than any model could”. This argument conflates general knowledge about a person with the specific knowledge needed to make a decision. Thus deciding whether or not to order an appendectomy after complaints of severe stomach pains, or do a brain scan after a head injury, should likely be based on a very specific set of criteria. Thus, a clinician’s sense of superiority about being more knowledgeable about a broader set of features that relate to the patient may be misplaced. 

Grove and Meehl \cite{grove1996comparative} gave the example of how physicians are sometimes more willing to accept risk than patients are because they feel that their knowledge of individual cases over-rules the general statistical relationships that apply in a set of data. A patient may have an annoying medical condition but be unwilling to accept a 5\% risk of dying during an operation to treat the condition. The surgeon on the other hand may feel like his knowledge of the case indicates that the risk is lower. "Personalizing" expert assessments in this way may often be wrong due to cognitive biases and heuristics such as the availability and representativeness heuristics \cite{tversky1974judgment}.

\section{Conclusion}
\label{sec:conclusion}

The “passing of the torch” between generations is often challenging. A father may be unwilling to continue playing after he has taught his child to play chess only to find that he now loses every game. Governments may feel the need to regulate the use of AI models to minimize job losses, and they will need to ensure that there is accountability, transparency, and fairness in the deployment of those models.

Progress towards model mastery is occurring across a wide range of applications. We need to recognize where AI systems are in the model mastery lifecycle for various applications and design task allocations and HAII accordingly. There should always be a degree of healthy skepticism when considering the possibility of model mastery. Models should be tested for brittleness. Are there edge cases that they do not handle well? Models should also be tested for sensitivity to drift. Is it possible that the characteristics of the decision problem are changing over time so that a model that was previously successful may no longer fit the data as well? 

We can use a statistical analogy to characterize the dilemma of when to accept Model Mastery and give AI models relative autonomy. If we implement model autonomy but overall task performance worsens, we have committed a Type 1 error where we rejected the status quo where the human remains “in-the-loop”. On the other hand, if we fail to implement model autonomy for a model that is demonstrably better and task performance suffers as a result, then we have committed a Type II error, accepting the status quo when we should have rejected it. In the traditional statistical approach you have to be very sure before rejecting the status quo (null hypothesis) and this level of caution is likely required in the design of AI systems as well. Just as a pilot has to earn her wings, so too AI models should prove that they have achieved mastery and can act more autonomously. 

While we advise caution in accepting model mastery, based on objective analyses of task performance, we should also be cautious about accepting persuasive arguments against model mastery that are not backed up by data, since efforts to avoid Type I errors should not lead to too many Type II errors. 

Model mastery seems easier to accept when the model is creating new capabilities that were not previously available, e.g. leveraging big data to speed up the process of drug discovery. However, when model mastery occurs in a domain that has been dominated by skilled human experts, then the implementation of successful models becomes more problematic. Although model mastery appears to be relatively prevalent in clinical decision making tasks \cite{grove1996comparative}, uptake of automated methods has been relatively slow. One reason for slow uptake of automated clinical decision making may be the view that clinicians have privileged expertise and are better aware of the decision making context. In this view, it might be better for AI to make suggestions and perhaps physician decisions can be improved through feedback. However, studies that have looked at this issue have not yielded encouraging results. For instance, Goldberg \cite{goldberg1965diagnosticians} gave judges immediate feedback on the accuracy of their judgments. In spite of the feedback provided, the clinicians were outperformed by a four-variable equally weighted regression equation. 

As AI models progress towards mastery across a wide range of domains we need more rational allocation of human/machine expertise where HAII appropriate to the level of progress towards model mastery can be implemented, and with sufficient safeguards against factors like brittleness, model drift, and violations of ethics and fairness. 

In conclusion, position on the model mastery lifecycle is an important factor in the design of HAII. While AI may outperform humans in specific tasks, human engagement still remains crucial in the vast majority of applications and thus HAII is also crucial. Our model mastery lifecycle framework seeks to provide guidance for integrating AI systems into work practices while accommodating the changing requirements for HAII as model expertise increases. Healthcare is a domain of particular interest and concern as many applications appear to be in or approaching the zone of uncertainty and because decisions have significant ethical and safety implications. While model mastery is an important issue, there will be many tasks where human judgment, advice, and interpretation will remain necessary and where HAII will be a critical determinant of overall system performance.

\bibliographystyle{ACM-Reference-Format}
\bibliography{main}


\begin{thebibliography}{00}

%%% the Web address.

\ifx \showCODEN    \undefined \def \showCODEN     #1{\unskip}     \fi
\ifx \showDOI      \undefined \def \showDOI       #1{#1}\fi
\ifx \showISBNx    \undefined \def \showISBNx     #1{\unskip}     \fi
\ifx \showISBNxiii \undefined \def \showISBNxiii  #1{\unskip}     \fi
\ifx \showISSN     \undefined \def \showISSN      #1{\unskip}     \fi
\ifx \showLCCN     \undefined \def \showLCCN      #1{\unskip}     \fi
\ifx \shownote     \undefined \def \shownote      #1{#1}          \fi
\ifx \showarticletitle \undefined \def \showarticletitle #1{#1}   \fi
\ifx \showURL      \undefined \def \showURL       {\relax}        \fi
% invisible to TeX
\providecommand\bibfield[2]{#2}
\providecommand\bibinfo[2]{#2}
\providecommand\natexlab[1]{#1}
\providecommand\showeprint[2][]{arXiv:#2}

\bibitem[\protect\citeauthoryear{Adam, Diebel, Goutier, and Benlian}{Adam et~al\mbox{.}}{2024}]%
        {adam2024navigating}
\bibfield{author}{\bibinfo{person}{Martin Adam}, \bibinfo{person}{Christopher Diebel}, \bibinfo{person}{Marc Goutier}, {and} \bibinfo{person}{Alexander Benlian}.} \bibinfo{year}{2024}\natexlab{}.
\newblock \showarticletitle{Navigating autonomy and control in human-AI delegation: User responses to technology-versus user-invoked task allocation}.
\newblock \bibinfo{journal}{{\em Decision Support Systems\/}} (\bibinfo{year}{2024}), \bibinfo{pages}{114193}.
\newblock


\bibitem[\protect\citeauthoryear{Amershi, Weld, Vorvoreanu, Fourney, Nushi, Collisson, Suh, Iqbal, Bennett, Inkpen, et~al\mbox{.}}{Amershi et~al\mbox{.}}{2019}]%
        {amershi2019guidelines}
\bibfield{author}{\bibinfo{person}{Saleema Amershi}, \bibinfo{person}{Dan Weld}, \bibinfo{person}{Mihaela Vorvoreanu}, \bibinfo{person}{Adam Fourney}, \bibinfo{person}{Besmira Nushi}, \bibinfo{person}{Penny Collisson}, \bibinfo{person}{Jina Suh}, \bibinfo{person}{Shamsi Iqbal}, \bibinfo{person}{Paul~N Bennett}, \bibinfo{person}{Kori Inkpen}, {et~al\mbox{.}}} \bibinfo{year}{2019}\natexlab{}.
\newblock \showarticletitle{Guidelines for human-AI interaction}. In \bibinfo{booktitle}{{\em Proceedings of the 2019 chi conference on human factors in computing systems}}. \bibinfo{pages}{1--13}.
\newblock


\bibitem[\protect\citeauthoryear{Cai, Winter, Steiner, Wilcox, and Terry}{Cai et~al\mbox{.}}{2019}]%
        {cai2019hello}
\bibfield{author}{\bibinfo{person}{Carrie~J Cai}, \bibinfo{person}{Samantha Winter}, \bibinfo{person}{David Steiner}, \bibinfo{person}{Lauren Wilcox}, {and} \bibinfo{person}{Michael Terry}.} \bibinfo{year}{2019}\natexlab{}.
\newblock \showarticletitle{" Hello AI": uncovering the onboarding needs of medical practitioners for human-AI collaborative decision-making}.
\newblock \bibinfo{journal}{{\em Proceedings of the ACM on Human-computer Interaction\/}} \bibinfo{volume}{3}, \bibinfo{number}{CSCW} (\bibinfo{year}{2019}), \bibinfo{pages}{1--24}.
\newblock


\bibitem[\protect\citeauthoryear{Campbell, Hoane~Jr, and Hsu}{Campbell et~al\mbox{.}}{2002}]%
        {campbell2002deep}
\bibfield{author}{\bibinfo{person}{Murray Campbell}, \bibinfo{person}{A~Joseph Hoane~Jr}, {and} \bibinfo{person}{Feng-hsiung Hsu}.} \bibinfo{year}{2002}\natexlab{}.
\newblock \showarticletitle{Deep blue}.
\newblock \bibinfo{journal}{{\em Artificial intelligence\/}} \bibinfo{volume}{134}, \bibinfo{number}{1-2} (\bibinfo{year}{2002}), \bibinfo{pages}{57--83}.
\newblock


\bibitem[\protect\citeauthoryear{Chignell, Wang, Zare, and Li}{Chignell et~al\mbox{.}}{2023}]%
        {chignell2023evolution}
\bibfield{author}{\bibinfo{person}{Mark Chignell}, \bibinfo{person}{Lu Wang}, \bibinfo{person}{Atefeh Zare}, {and} \bibinfo{person}{Jamy Li}.} \bibinfo{year}{2023}\natexlab{}.
\newblock \showarticletitle{The evolution of HCI and human factors: Integrating human and artificial intelligence}.
\newblock \bibinfo{journal}{{\em ACM Transactions on Computer-Human Interaction\/}} \bibinfo{volume}{30}, \bibinfo{number}{2} (\bibinfo{year}{2023}), \bibinfo{pages}{1--30}.
\newblock


\bibitem[\protect\citeauthoryear{Chung, Wang, Li, Yang, Giang, Jerath, Raman, Lie, and Chignell}{Chung et~al\mbox{.}}{2023}]%
        {chung2023implementing}
\bibfield{author}{\bibinfo{person}{Mu-Huan Chung}, \bibinfo{person}{Lu Wang}, \bibinfo{person}{Sharon Li}, \bibinfo{person}{Yuhong Yang}, \bibinfo{person}{Calvin Giang}, \bibinfo{person}{Khilan Jerath}, \bibinfo{person}{Abhay Raman}, \bibinfo{person}{David Lie}, {and} \bibinfo{person}{Mark Chignell}.} \bibinfo{year}{2023}\natexlab{}.
\newblock \showarticletitle{Implementing Active Learning in Cybersecurity: Detecting Anomalies in Redacted Emails}.
\newblock \bibinfo{journal}{{\em arXiv e-prints\/}} (\bibinfo{year}{2023}), \bibinfo{pages}{arXiv--2303}.
\newblock


\bibitem[\protect\citeauthoryear{Dawes}{Dawes}{2005}]%
        {dawes2005ethical}
\bibfield{author}{\bibinfo{person}{Robyn~M Dawes}.} \bibinfo{year}{2005}\natexlab{}.
\newblock \showarticletitle{The ethical implications of Paul Meehl's work on comparing clinical versus actuarial prediction methods}.
\newblock \bibinfo{journal}{{\em Journal of Clinical Psychology\/}} \bibinfo{volume}{61}, \bibinfo{number}{10} (\bibinfo{year}{2005}), \bibinfo{pages}{1245--1255}.
\newblock


\bibitem[\protect\citeauthoryear{Dawes and Corrigan}{Dawes and Corrigan}{1974}]%
        {dawes1974linear}
\bibfield{author}{\bibinfo{person}{Robyn~M Dawes} {and} \bibinfo{person}{Bernard Corrigan}.} \bibinfo{year}{1974}\natexlab{}.
\newblock \showarticletitle{Linear models in decision making.}
\newblock \bibinfo{journal}{{\em Psychological bulletin\/}} \bibinfo{volume}{81}, \bibinfo{number}{2} (\bibinfo{year}{1974}), \bibinfo{pages}{95}.
\newblock


\bibitem[\protect\citeauthoryear{Elkins and Chun}{Elkins and Chun}{2020}]%
        {elkins2020can}
\bibfield{author}{\bibinfo{person}{Katherine Elkins} {and} \bibinfo{person}{Jon Chun}.} \bibinfo{year}{2020}\natexlab{}.
\newblock \showarticletitle{Can GPT-3 pass a writer’s Turing test?}
\newblock \bibinfo{journal}{{\em Journal of Cultural Analytics\/}} \bibinfo{volume}{5}, \bibinfo{number}{2} (\bibinfo{year}{2020}).
\newblock


\bibitem[\protect\citeauthoryear{Froomkin, Kerr, and Pineau}{Froomkin et~al\mbox{.}}{2019}]%
        {froomkin2019ais}
\bibfield{author}{\bibinfo{person}{A~Michael Froomkin}, \bibinfo{person}{Ian Kerr}, {and} \bibinfo{person}{Joelle Pineau}.} \bibinfo{year}{2019}\natexlab{}.
\newblock \showarticletitle{When AIs outperform doctors: confronting the challenges of a tort-induced over-reliance on machine learning}.
\newblock \bibinfo{journal}{{\em Ariz. L. Rev.\/}}  \bibinfo{volume}{61} (\bibinfo{year}{2019}), \bibinfo{pages}{33}.
\newblock


\bibitem[\protect\citeauthoryear{Gama, Tyskbo, Nygren, Barlow, Reed, and Svedberg}{Gama et~al\mbox{.}}{2022}]%
        {gama2022implementation}
\bibfield{author}{\bibinfo{person}{F{\'a}bio Gama}, \bibinfo{person}{Daniel Tyskbo}, \bibinfo{person}{Jens Nygren}, \bibinfo{person}{James Barlow}, \bibinfo{person}{Julie Reed}, {and} \bibinfo{person}{Petra Svedberg}.} \bibinfo{year}{2022}\natexlab{}.
\newblock \showarticletitle{Implementation frameworks for artificial intelligence translation into health care practice: scoping review}.
\newblock \bibinfo{journal}{{\em Journal of medical Internet research\/}} \bibinfo{volume}{24}, \bibinfo{number}{1} (\bibinfo{year}{2022}), \bibinfo{pages}{e32215}.
\newblock


\bibitem[\protect\citeauthoryear{Goldberg}{Goldberg}{1965}]%
        {goldberg1965diagnosticians}
\bibfield{author}{\bibinfo{person}{Lewis~R Goldberg}.} \bibinfo{year}{1965}\natexlab{}.
\newblock \showarticletitle{Diagnosticians vs. diagnostic signs: The diagnosis of psychosis vs. neurosis from the MMPI.}
\newblock \bibinfo{journal}{{\em Psychological Monographs: General and Applied\/}} \bibinfo{volume}{79}, \bibinfo{number}{9} (\bibinfo{year}{1965}), \bibinfo{pages}{1}.
\newblock


\bibitem[\protect\citeauthoryear{Grove and Meehl}{Grove and Meehl}{1996}]%
        {grove1996comparative}
\bibfield{author}{\bibinfo{person}{William~M Grove} {and} \bibinfo{person}{Paul~E Meehl}.} \bibinfo{year}{1996}\natexlab{}.
\newblock \showarticletitle{Comparative efficiency of informal (subjective, impressionistic) and formal (mechanical, algorithmic) prediction procedures: The clinical--statistical controversy.}
\newblock \bibinfo{journal}{{\em Psychology, public policy, and law\/}} \bibinfo{volume}{2}, \bibinfo{number}{2} (\bibinfo{year}{1996}), \bibinfo{pages}{293}.
\newblock


\bibitem[\protect\citeauthoryear{Hancock, Kajaks, Caird, Chignell, Mizobuchi, Burns, Feng, Fernie, Lavalli{\`e}re, Noy, et~al\mbox{.}}{Hancock et~al\mbox{.}}{2020}]%
        {hancock2020challenges}
\bibfield{author}{\bibinfo{person}{Peter~A Hancock}, \bibinfo{person}{Tara Kajaks}, \bibinfo{person}{Jeff~K Caird}, \bibinfo{person}{Mark~H Chignell}, \bibinfo{person}{Sachi Mizobuchi}, \bibinfo{person}{Peter~C Burns}, \bibinfo{person}{Jing Feng}, \bibinfo{person}{Geoff~R Fernie}, \bibinfo{person}{Martin Lavalli{\`e}re}, \bibinfo{person}{Ian~Y Noy}, {et~al\mbox{.}}} \bibinfo{year}{2020}\natexlab{}.
\newblock \showarticletitle{Challenges to human drivers in increasingly automated vehicles}.
\newblock \bibinfo{journal}{{\em Human factors\/}} \bibinfo{volume}{62}, \bibinfo{number}{2} (\bibinfo{year}{2020}), \bibinfo{pages}{310--328}.
\newblock


\bibitem[\protect\citeauthoryear{Huang and Rust}{Huang and Rust}{2021}]%
        {huang2021engaged}
\bibfield{author}{\bibinfo{person}{Ming-Hui Huang} {and} \bibinfo{person}{Roland~T Rust}.} \bibinfo{year}{2021}\natexlab{}.
\newblock \showarticletitle{Engaged to a robot? The role of AI in service}.
\newblock \bibinfo{journal}{{\em Journal of Service Research\/}} \bibinfo{volume}{24}, \bibinfo{number}{1} (\bibinfo{year}{2021}), \bibinfo{pages}{30--41}.
\newblock


\bibitem[\protect\citeauthoryear{Kochenderfer, Chryssanthacopoulos, and Weibel}{Kochenderfer et~al\mbox{.}}{2012}]%
        {kochenderfer2012new}
\bibfield{author}{\bibinfo{person}{Mykel~J Kochenderfer}, \bibinfo{person}{James~P Chryssanthacopoulos}, {and} \bibinfo{person}{Roland~E Weibel}.} \bibinfo{year}{2012}\natexlab{}.
\newblock \showarticletitle{A new approach for designing safer collision avoidance systems}.
\newblock \bibinfo{journal}{{\em Air Traffic Control Quarterly\/}} \bibinfo{volume}{20}, \bibinfo{number}{1} (\bibinfo{year}{2012}), \bibinfo{pages}{27--45}.
\newblock


\bibitem[\protect\citeauthoryear{Kosch, Welsch, Chuang, and Schmidt}{Kosch et~al\mbox{.}}{2023}]%
        {kosch2023placebo}
\bibfield{author}{\bibinfo{person}{Thomas Kosch}, \bibinfo{person}{Robin Welsch}, \bibinfo{person}{Lewis Chuang}, {and} \bibinfo{person}{Albrecht Schmidt}.} \bibinfo{year}{2023}\natexlab{}.
\newblock \showarticletitle{The placebo effect of artificial intelligence in human--computer interaction}.
\newblock \bibinfo{journal}{{\em ACM Transactions on Computer-Human Interaction\/}} \bibinfo{volume}{29}, \bibinfo{number}{6} (\bibinfo{year}{2023}), \bibinfo{pages}{1--32}.
\newblock


\bibitem[\protect\citeauthoryear{Kruger and Dunning}{Kruger and Dunning}{1999}]%
        {kruger1999unskilled}
\bibfield{author}{\bibinfo{person}{Justin Kruger} {and} \bibinfo{person}{David Dunning}.} \bibinfo{year}{1999}\natexlab{}.
\newblock \showarticletitle{Unskilled and unaware of it: how difficulties in recognizing one's own incompetence lead to inflated self-assessments.}
\newblock \bibinfo{journal}{{\em Journal of personality and social psychology\/}} \bibinfo{volume}{77}, \bibinfo{number}{6} (\bibinfo{year}{1999}), \bibinfo{pages}{1121}.
\newblock


\bibitem[\protect\citeauthoryear{Kurzweil}{Kurzweil}{2000}]%
        {kurzweil2000age}
\bibfield{author}{\bibinfo{person}{Ray Kurzweil}.} \bibinfo{year}{2000}\natexlab{}.
\newblock \bibinfo{booktitle}{{\em The age of spiritual machines: When computers exceed human intelligence}}.
\newblock \bibinfo{publisher}{Penguin}.
\newblock


\bibitem[\protect\citeauthoryear{Lai, Chen, Smith-Renner, Liao, and Tan}{Lai et~al\mbox{.}}{2023}]%
        {lai2023towards}
\bibfield{author}{\bibinfo{person}{Vivian Lai}, \bibinfo{person}{Chacha Chen}, \bibinfo{person}{Alison Smith-Renner}, \bibinfo{person}{Q~Vera Liao}, {and} \bibinfo{person}{Chenhao Tan}.} \bibinfo{year}{2023}\natexlab{}.
\newblock \showarticletitle{Towards a science of human-ai decision making: An overview of design space in empirical human-subject studies}. In \bibinfo{booktitle}{{\em Proceedings of the 2023 ACM Conference on Fairness, Accountability, and Transparency}}. \bibinfo{pages}{1369--1385}.
\newblock


\bibitem[\protect\citeauthoryear{Li, Vorvoreanu, DeBellis, and Amershi}{Li et~al\mbox{.}}{2023}]%
        {li2023assessing}
\bibfield{author}{\bibinfo{person}{Tianyi Li}, \bibinfo{person}{Mihaela Vorvoreanu}, \bibinfo{person}{Derek DeBellis}, {and} \bibinfo{person}{Saleema Amershi}.} \bibinfo{year}{2023}\natexlab{}.
\newblock \showarticletitle{Assessing human-ai interaction early through factorial surveys: A study on the guidelines for human-ai interaction}.
\newblock \bibinfo{journal}{{\em ACM Transactions on Computer-Human Interaction\/}} \bibinfo{volume}{30}, \bibinfo{number}{5} (\bibinfo{year}{2023}), \bibinfo{pages}{1--45}.
\newblock


\bibitem[\protect\citeauthoryear{Logg, Minson, and Moore}{Logg et~al\mbox{.}}{2019}]%
        {logg2019algorithm}
\bibfield{author}{\bibinfo{person}{Jennifer~M Logg}, \bibinfo{person}{Julia~A Minson}, {and} \bibinfo{person}{Don~A Moore}.} \bibinfo{year}{2019}\natexlab{}.
\newblock \showarticletitle{Algorithm appreciation: People prefer algorithmic to human judgment}.
\newblock \bibinfo{journal}{{\em Organizational Behavior and Human Decision Processes\/}}  \bibinfo{volume}{151} (\bibinfo{year}{2019}), \bibinfo{pages}{90--103}.
\newblock


\bibitem[\protect\citeauthoryear{Mahmud, Hong, and Fong}{Mahmud et~al\mbox{.}}{2023}]%
        {mahmud2023study}
\bibfield{author}{\bibinfo{person}{Bahar Mahmud}, \bibinfo{person}{Guan Hong}, {and} \bibinfo{person}{Bernard Fong}.} \bibinfo{year}{2023}\natexlab{}.
\newblock \showarticletitle{A Study of Human--AI Symbiosis for Creative Work: Recent Developments and Future Directions in Deep Learning}.
\newblock \bibinfo{journal}{{\em ACM Transactions on Multimedia Computing, Communications and Applications\/}} \bibinfo{volume}{20}, \bibinfo{number}{2} (\bibinfo{year}{2023}), \bibinfo{pages}{1--21}.
\newblock


\bibitem[\protect\citeauthoryear{Meehl}{Meehl}{1954}]%
        {meehl1954clinical}
\bibfield{author}{\bibinfo{person}{Paul~E Meehl}.} \bibinfo{year}{1954}\natexlab{}.
\newblock \showarticletitle{Clinical versus statistical prediction: A theoretical analysis and a review of the evidence.}
\newblock  (\bibinfo{year}{1954}).
\newblock


\bibitem[\protect\citeauthoryear{Oberije, Nalbantov, Dekker, Boersma, Borger, Reymen, van Baardwijk, Wanders, De~Ruysscher, Steyerberg, et~al\mbox{.}}{Oberije et~al\mbox{.}}{2014}]%
        {oberije2014prospective}
\bibfield{author}{\bibinfo{person}{Cary Oberije}, \bibinfo{person}{Georgi Nalbantov}, \bibinfo{person}{Andre Dekker}, \bibinfo{person}{Liesbeth Boersma}, \bibinfo{person}{Jacques Borger}, \bibinfo{person}{Bart Reymen}, \bibinfo{person}{Angela van Baardwijk}, \bibinfo{person}{Rinus Wanders}, \bibinfo{person}{Dirk De~Ruysscher}, \bibinfo{person}{Ewout Steyerberg}, {et~al\mbox{.}}} \bibinfo{year}{2014}\natexlab{}.
\newblock \showarticletitle{A prospective study comparing the predictions of doctors versus models for treatment outcome of lung cancer patients: a step toward individualized care and shared decision making}.
\newblock \bibinfo{journal}{{\em Radiotherapy and Oncology\/}} \bibinfo{volume}{112}, \bibinfo{number}{1} (\bibinfo{year}{2014}), \bibinfo{pages}{37--43}.
\newblock


\bibitem[\protect\citeauthoryear{of~T~News}{of~T~News}{2023}]%
        {UofTNews}
\bibfield{author}{\bibinfo{person}{U of T~News}.} \bibinfo{year}{2023}\natexlab{}.
\newblock \bibinfo{booktitle}{{\em The Godfather in Conversation: Why Geoffrey Hinton is worried about the future of AI}}.
\newblock \bibinfo{type}{{T}echnical {R}eport}. \bibinfo{institution}{University of Toronto}.
\newblock


\bibitem[\protect\citeauthoryear{Sellen and Horvitz}{Sellen and Horvitz}{2023}]%
        {sellen2023rise}
\bibfield{author}{\bibinfo{person}{Abigail Sellen} {and} \bibinfo{person}{Eric Horvitz}.} \bibinfo{year}{2023}\natexlab{}.
\newblock \showarticletitle{The rise of the AI Co-Pilot: Lessons for design from aviation and beyond}.
\newblock \bibinfo{journal}{{\em arXiv preprint arXiv:2311.14713\/}} (\bibinfo{year}{2023}).
\newblock


\bibitem[\protect\citeauthoryear{Settles}{Settles}{2011}]%
        {settles2011theories}
\bibfield{author}{\bibinfo{person}{Burr Settles}.} \bibinfo{year}{2011}\natexlab{}.
\newblock \showarticletitle{From theories to queries: Active learning in practice}. In \bibinfo{booktitle}{{\em Active learning and experimental design workshop in conjunction with AISTATS 2010}}. JMLR Workshop and Conference Proceedings, \bibinfo{pages}{1--18}.
\newblock


\bibitem[\protect\citeauthoryear{Sharma, Lin, Miner, Atkins, and Althoff}{Sharma et~al\mbox{.}}{2023}]%
        {sharma2023human}
\bibfield{author}{\bibinfo{person}{Ashish Sharma}, \bibinfo{person}{Inna~W Lin}, \bibinfo{person}{Adam~S Miner}, \bibinfo{person}{David~C Atkins}, {and} \bibinfo{person}{Tim Althoff}.} \bibinfo{year}{2023}\natexlab{}.
\newblock \showarticletitle{Human--AI collaboration enables more empathic conversations in text-based peer-to-peer mental health support}.
\newblock \bibinfo{journal}{{\em Nature Machine Intelligence\/}} \bibinfo{volume}{5}, \bibinfo{number}{1} (\bibinfo{year}{2023}), \bibinfo{pages}{46--57}.
\newblock


\bibitem[\protect\citeauthoryear{Sheridan}{Sheridan}{2012}]%
        {sheridan2012human}
\bibfield{author}{\bibinfo{person}{Thomas~B Sheridan}.} \bibinfo{year}{2012}\natexlab{}.
\newblock \showarticletitle{Human supervisory control}.
\newblock \bibinfo{journal}{{\em Handbook of human factors and ergonomics\/}} (\bibinfo{year}{2012}), \bibinfo{pages}{990--1015}.
\newblock


\bibitem[\protect\citeauthoryear{Shneiderman}{Shneiderman}{2022}]%
        {shneiderman2022human}
\bibfield{author}{\bibinfo{person}{Ben Shneiderman}.} \bibinfo{year}{2022}\natexlab{}.
\newblock \bibinfo{booktitle}{{\em Human-centered AI}}.
\newblock \bibinfo{publisher}{Oxford University Press}.
\newblock


\bibitem[\protect\citeauthoryear{Shuaib, Arian, and Shuaib}{Shuaib et~al\mbox{.}}{2020}]%
        {shuaib2020increasing}
\bibfield{author}{\bibinfo{person}{Abdullah Shuaib}, \bibinfo{person}{Husain Arian}, {and} \bibinfo{person}{Ali Shuaib}.} \bibinfo{year}{2020}\natexlab{}.
\newblock \showarticletitle{The increasing role of artificial intelligence in health care: Will robots replace doctors in the future?}
\newblock \bibinfo{journal}{{\em International journal of general medicine\/}} (\bibinfo{year}{2020}), \bibinfo{pages}{891--896}.
\newblock


\bibitem[\protect\citeauthoryear{Skraaning and Jamieson}{Skraaning and Jamieson}{2021}]%
        {skraaning2021human}
\bibfield{author}{\bibinfo{person}{Gyrd Skraaning} {and} \bibinfo{person}{Greg~A Jamieson}.} \bibinfo{year}{2021}\natexlab{}.
\newblock \showarticletitle{Human performance benefits of the automation transparency design principle: Validation and variation}.
\newblock \bibinfo{journal}{{\em Human factors\/}} \bibinfo{volume}{63}, \bibinfo{number}{3} (\bibinfo{year}{2021}), \bibinfo{pages}{379--401}.
\newblock


\bibitem[\protect\citeauthoryear{Thompson, Valdes, Fuller, Carpenter, Morin, Aneja, Lindsay, Aerts, Agrimson, Deville~Jr, et~al\mbox{.}}{Thompson et~al\mbox{.}}{2018}]%
        {thompson2018artificial}
\bibfield{author}{\bibinfo{person}{Reid~F Thompson}, \bibinfo{person}{Gilmer Valdes}, \bibinfo{person}{Clifton~D Fuller}, \bibinfo{person}{Colin~M Carpenter}, \bibinfo{person}{Olivier Morin}, \bibinfo{person}{Sanjay Aneja}, \bibinfo{person}{William~D Lindsay}, \bibinfo{person}{Hugo~JWL Aerts}, \bibinfo{person}{Barbara Agrimson}, \bibinfo{person}{Curtiland Deville~Jr}, {et~al\mbox{.}}} \bibinfo{year}{2018}\natexlab{}.
\newblock \showarticletitle{Artificial intelligence in radiation oncology: a specialty-wide disruptive transformation?}
\newblock \bibinfo{journal}{{\em Radiotherapy and Oncology\/}} \bibinfo{volume}{129}, \bibinfo{number}{3} (\bibinfo{year}{2018}), \bibinfo{pages}{421--426}.
\newblock


\bibitem[\protect\citeauthoryear{Topol}{Topol}{2019}]%
        {topol2019high}
\bibfield{author}{\bibinfo{person}{Eric~J Topol}.} \bibinfo{year}{2019}\natexlab{}.
\newblock \showarticletitle{High-performance medicine: the convergence of human and artificial intelligence}.
\newblock \bibinfo{journal}{{\em Nature medicine\/}} \bibinfo{volume}{25}, \bibinfo{number}{1} (\bibinfo{year}{2019}), \bibinfo{pages}{44--56}.
\newblock


\bibitem[\protect\citeauthoryear{Tversky and Kahneman}{Tversky and Kahneman}{1974}]%
        {tversky1974judgment}
\bibfield{author}{\bibinfo{person}{Amos Tversky} {and} \bibinfo{person}{Daniel Kahneman}.} \bibinfo{year}{1974}\natexlab{}.
\newblock \showarticletitle{Judgment under Uncertainty: Heuristics and Biases: Biases in judgments reveal some heuristics of thinking under uncertainty.}
\newblock \bibinfo{journal}{{\em science\/}} \bibinfo{volume}{185}, \bibinfo{number}{4157} (\bibinfo{year}{1974}), \bibinfo{pages}{1124--1131}.
\newblock


\bibitem[\protect\citeauthoryear{Xu and Gao}{Xu and Gao}{2024}]%
        {xu2024applying}
\bibfield{author}{\bibinfo{person}{Wei Xu} {and} \bibinfo{person}{Zaifeng Gao}.} \bibinfo{year}{2024}\natexlab{}.
\newblock \showarticletitle{Applying HCAI in developing effective human-AI teaming: A perspective from human-AI joint cognitive systems}.
\newblock \bibinfo{journal}{{\em Interactions\/}} \bibinfo{volume}{31}, \bibinfo{number}{1} (\bibinfo{year}{2024}), \bibinfo{pages}{32--37}.
\newblock


\bibitem[\protect\citeauthoryear{Yildirim, Pushkarna, Goyal, Wattenberg, and Vi{\'e}gas}{Yildirim et~al\mbox{.}}{2023}]%
        {yildirim2023investigating}
\bibfield{author}{\bibinfo{person}{Nur Yildirim}, \bibinfo{person}{Mahima Pushkarna}, \bibinfo{person}{Nitesh Goyal}, \bibinfo{person}{Martin Wattenberg}, {and} \bibinfo{person}{Fernanda Vi{\'e}gas}.} \bibinfo{year}{2023}\natexlab{}.
\newblock \showarticletitle{Investigating how practitioners use human-ai guidelines: A case study on the people+ ai guidebook}. In \bibinfo{booktitle}{{\em Proceedings of the 2023 CHI Conference on Human Factors in Computing Systems}}. \bibinfo{pages}{1--13}.
\newblock


\end{thebibliography}

% % --- Appendix ---%

\end{document}